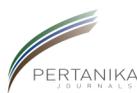

# Localization of Fourier-Laplace Series of Distributions

[1,2]Anvarjon Ahmedov, [2*]Ahmad Fadly Nurullah
and [3]Abdumalik Rakhimov

[1]*Department of Process and Food Engineering, Faculty of Engineering, Universiti Putra Malaysia, 43400 UPM Serdang, Selangor, Malaysia*

[2]*Institute for Mathematical Research, Universiti Putra Malaysia, 43400 UPM Serdang, Selangor, Malaysia*

[3]*Department of Science in Engineering, International Islamic University Malaysia, 50728 Kuala Lumpur, Malaysia*

*E-mail: anvar@eng.upm.edu.my and ahmadfadlynurullah@yahoo.com*

*Corresponding author

## ABSTRACT

This work was intended as an attempt to extend the results on localization of Fourier-Laplace series to the spectral expansions of distributions on the unit sphere. It is shown that the spectral expansions of the distribution on the unit sphere can be represented in terms of decompostions of Laplace-Beltrami operator. It was of interest to establish sufficient conditions for localization of the spectral expansions of distribution to clarify the latter some relevant counter examples are indicated.

## 1. INTRODUCTION

Let $S^N$ be unit sphere in $R^{N+1}$.

$$S^N = \{x \in R^{N+1} : |x|^2 = x_1^2 + x_2^2 + ... + x_{N+1}^2 = 1\}.$$

The sphere $S^N$ is naturally equipped with a positive measure $d\sigma(x)$ and with an elliptic second order differential operator $-\Delta_s$, namely the Laplace-Beltrami operator on the sphere. This operator is symmetric and nonnegative, and it can be extended to a nonnegative self-adjoint operator on the space $L_2(S^N)$. Where $L_2(S^N)$ denoted the $L_2$-space associated with the measure $d\sigma(x)$ on the sphere. For the self-adjoint extension of the Laplace-Beltrami operator we use again the same symbol $\Delta_s$ and by $\{\lambda_k\}, k = 0,1,2,...$ we denote the set of the eigenvalues of the Laplace-Beltrami operator $\Delta_s$, which is increasing sequence of nonnegative eigenvalues $\lambda_k = k(k+N-1), k = 0,1,2,...$ with finite multiplicities $a_0 = 1, a_1 = N, a_k = \frac{(N+k)!}{N!k!} - \frac{(N+k-1)!}{N!(k-2)!} \approx k^{N-1}, k \geq 2$ (and written as such) tending to infinity. We denote by $Y_j^k(x)$ the eigenfunctions of the Laplace-Beltrami operator corresponding to $\lambda_k$:

$$\Delta_s Y_j^{(k)}(x) = \lambda_k Y_j^{(k)}, \; j = 1,2,...,a_k; k = 0,1,2,... \; .$$

The system of eigenfunctions of the Laplace-Beltrami operator form an orthonormal basis in $L_2(S^N)$. To any measurable function $f$ we assign its spectral expansion:

$$f(x) = \sum_{k=0}^{\infty} Y_k(f,x),$$

where

$$Y_k(f,x) = \sum_{j=0}^{a_k} Y_j^{(k)}(x) \int_{S^N} f(y) Y_j^{(k)}(y) d\sigma(y), k = 0,1,2,...$$

The operator

$$E_\lambda f(x) = \int_{S^N} f(y) \Theta(x,y,\lambda) d\sigma(y)$$

is called spectral expansions of $f$, where

$$\Theta(x,y,\lambda) = \sum_{\lambda_k < \lambda} \sum_{j=1}^{a_k} Y_j^{(k)}(x) Y_j^{(k)}(y).$$

In this work we investigate spectral expansions related to distributions. We denote by $D(S^N) = C^\infty(S^N)$ the space of infinite times differentiable functions with the topology of uniformly convergence with Localization of Fourier-Laplace Series of Distributionsrespect to derivatives of all order. Let $D'(S^N)$ denotes the conjugate space of $D(S^N)$ i.e. the space of the linear functionals defined on $D(S^N)$ which are called distributions on sphere. From ellipticity of Laplace-Beltrami operator we obtain that $E_\lambda \varphi \in C^\infty(S^N)$ for all $\varphi \in C^\infty(S^N)$.

With the help of the latter remark we define:

**Definition**. The spectral expansions of the distribution $f \in D'(S^N)$ is distribution $E_\lambda f \in D'(S^N)$ defined as follows:

$$<E_\lambda f, \varphi> = <f, E_\lambda \varphi>,$$

for any $\varphi \in D(S^N)$.

The correctness of this definition can be established using the fact that the operator $E_\lambda$ is linear and bounded from $D(S^N)$ to $D'(S^N)$. Furthermore we have

**Theorem 1.1** *Let $\varphi \in D(S^N)$, then the partial sums of the Fourier-Laplace series of the $\varphi$ converges in the topology of the space $D(S^N)$*

$$E_n \varphi(x) \to \varphi(x).$$

From here we obtain the following

**Theorem 1.2** *Let $f \in D'(S^N)$, then the partial sums of the Fourier-Laplace series of the $f$ converges in the topology of the space $D'(S^N)$*

$$<E_n f(x), g(x)> \to <f(x), g(x)>, \forall g \in D(S^N).$$

Our aim is to prove convergence of $E_\lambda f$ in classical means, but it can be proved only in domains where $f$ coincides with local integrable functions.

*Malaysian Journal of Mathematical Sciences*  317

For any distribution $f \in D'(S^N)$ we assign its Riesz means of the spectral expansions of $f$ as follows

$$E_n^\alpha f(x) = <f, \Theta^\alpha(x,y,n)> = \sum_{k=0}^{n}\left(1-\frac{\lambda_k}{\lambda_n}\right)_+^\alpha \sum_{j=1}^{a_k} Y_j^k(x) <f, Y_j^k(y)>,$$

where $<f, Y_j^k(y)>$ is the value of the functional $f$ on the eigenfunction $Y_j^k(y)$ of the Laplace-Beltrami operator.

Ashurov and Anvarjon (2011) considered the spectral expansions of functions from Nikol'skii classes $H_p^a(R^N)$ related to selfadjoint extensions of elliptic differential operators A(D) of order $m$ in $R^N$. They then constructed a continuous function from Nikol'skii class with $p \cdot a > N$ such that the Riesz means of the spectral expansion diverge at the origin. This result demonstrates the sharpness of the $p \cdot a > N$ condition by Alimov (1976) for uniform convergence of spectral expansions, related to elliptic differential operators.

In 1983 Meaney (1984) presented a localization theorem for expansions in eigenfunctions of the Laplace–Beltrami operator on a compact two-point homogeneous space. When $X = S^1$ the theorem proved is just the usual localization principle for pseudo functions. The proof of his theorem mimics that of the one-dimensional case. In his paper, he manages to avoid the issue of localization of summability methods, as treated in Bonami and Clerc (1973).

Results in the work of Zhizhiashvili and Topuriya (1979) discuss the theory of Fourier-Laplace series on a N-dimensional sphere where $N \geq 3$. The theory of the Fourier-Laplace series has been developed intensively in various directions. From the entire variety of problems investigated in this field, the woks in Zhizhiashvili and Topuriya (1979) restrict the presentation of the theory of these series on a N-dimensional sphere $N \geq 3$.

Anvarjon, Norashikin in Anvarjon and Norashikin (2012) proved a localization theorem for nonspherical partial sums, that is, for Fourier series under summation over domains bounded by level surfaces of elliptic polynomials. The result provides a natural and intrinsic characterization of the approximation of the functions by Fourier–Laplace series.

For the latest investigations related to the convergence problems of the spectral expansions of the elliptic differential operators we refer the readers to Ashurov and Anvarjon (2010), Ashurov *et al.* (2010), Anvarjon (2010), Alimov (2006) and what we personally believe that provides a more complete insight is Alimov *et al.* (1992).

We define the Sobolev space $H_p^s(S^N)$ as follows: for $1 < p < \infty$, $s > 0$, $H_p^s(S^N)$ is the collection of all $f \in L_p(S^N)$ such that $f = (I - \Delta_s)^{-\frac{s}{2}} h$ for some $h \in L_p(S^N)$ with the norm

$$\|f\|_{H_p^s(S^N)} = \|h\|_{L_p(S^N)}.$$

If $s < 0$; $H_p^s(S^N)$ is the collection of all $f \in D'(S^N)$ of the form $f = (I - \Delta_s)^k h$ with $h \in H_p^{2k+s}(S^N)$, where $k$ is a natural number such that $2k + s > 0$ and

$$\|f\|_{H_p^s(S^N)} = \|h\|_{H_p^{2k+s}(S^N)}.$$

If $s = 0$, $H_p^0(S^N) = L_p(S^N)$.

If $p = 2$, we have

$$\|f\|_{H_2^s(S^N)} = \left\| \sum_{k=0}^{\infty} (1 + \lambda_k)^{\frac{l}{2}} \sum_{j=1}^{a_k} <f, Y_j^k(\cdot)> Y_j^k(x) \right\|_{L_2(S^N)}.$$

It is important to remark here that the Nikolskii classes $H_2^l(S^N)$ are Hilbert spaces and Schwartz space $D(S^N)$ is dense in $H_2^l(S^N)$, $l > 0$. The equivalent definition of the norm in Nikolskii spaces $H_2^{-l}(S^N)$, $l > 0$ may be given by the following

$$\|f\|_{H_2^{-l}(S^N)} = \sup_{0 \neq u \in H_2^l(S^N)} \frac{\|<f, u>\|}{\|u\|_{H_2^l(S^N)}} \qquad (2)$$

## 2. MAIN RESULTS

The main results of the work are the following:

**Theorem 2.1** *Let $f \in H_2^{-l}$, $l > 0$ and $f$ vanishes in some domain $V \subset S^N$. If $\alpha \geq l + \frac{N-1}{2}$, then the Riesz means $E_n^\alpha f(x)$ uniformly converges to $0$ on any compact $K \subset V$.*

In the case when distribution coincides with normal function we have,

**Theorem 2.2** *Let a distribution $f \in H_2^{-l}$, $l > 0$ coincides with the continuous function $g(x)$ in some domain $V \subset S^N$. If $\alpha \geq l + \frac{N-1}{2}$ then*

$$\lim_{n \to \infty} E_n^\alpha f(x) = g(x)$$

*in any compact $K \subset V$.*

**Proof of Main Results**

The proof of the main facts are based on the following Lemmas.

**Lemma 2.3** *Let $V \subset S^N$ be a domain and $K \subset V$ be any compact set. Then uniformly on $x \in K$ one has*

$$\left\| \Theta^\alpha(x, y, n) \right\|_{L_2(S^N \setminus V)} \leq C n^{\frac{N-1}{2} - \alpha}.$$

**Proof.** The proof is straightfoward. From $(S^N \setminus V) \cap K = \emptyset$ we conclude that there exist $r_0 > 0$ such that, for any $y \in S^N \setminus V$ and $x \in K$ one has $\gamma(x, y) \geq r_0$. Based on the later we divide the integration region $S^N \setminus V$ into two parts as follows:

$$\left( \int_{S^N \setminus V} \left| \Theta^\alpha(x, y, n) \right|^2 d\sigma(y) \right)^{\frac{1}{2}} \leq$$

$$\leq \left( \int_{r_0 \leq \gamma \leq \pi - \frac{1}{n}} \left| \Theta^\alpha(x, y, n) \right|^2 d\sigma(y) \right)^{\frac{1}{2}} + \left( \int_{\pi - \frac{1}{n} \leq \gamma \leq \pi} \left| \Theta^\alpha(x, y, n) \right|^2 d\sigma(y) \right)^{\frac{1}{2}} = I_1 + I_2.$$

To estimate the $I_1$ and $I_2$ we use the following:

**Lemma 2.4** *Let $\Theta^\alpha(x,y,n)$, be the kernel of Riesz means of the spectral expansions*

1. If $\left\|\frac{\pi}{2}-\gamma\right\| < \frac{n}{n+1}\frac{\pi}{2}$, $n \to \infty$, then one has

$$\Theta^\alpha(x,y,n) = O(1)\left(\frac{n^{(N-1)/2}}{(\sin\gamma)^{(N-1)/2}(\sin(\gamma/2))^{1+\alpha}} + \frac{n^{(N-3)/2}}{(\sin\gamma)^{(N+1)/2}(\sin(\gamma/2))^{1+\alpha}}\right.$$

$$\left. + \frac{n^{-1}}{((\sin\gamma)/2)^{1+N}}\right);$$

2. if $0 \leq \gamma \leq \pi$, $n > 1$ then one has

$$\Theta^\alpha(x,y,n) = O(1)n^N,$$

3. if $0 \leq \gamma_0 \leq \gamma \leq \pi$, $n > 1$ then one has

$$\Theta^\alpha(x,y,n) = O(1)n^{N-1-\alpha},$$

for the proof see Anvarjon (2011).

Using this Lemma we first estimate $I_1$ as follows

$$I_1 \leq c_1 n^{\frac{N-1}{2}-\alpha}\sqrt{\int_{r_0 \leq \gamma \leq \pi - \frac{1}{n}}(\sin\gamma)^{-N+1}d\sigma(y)} + c_2 n^{\frac{N-3}{2}-\alpha}\sqrt{\int_{r_0 \leq \gamma \leq \pi - \frac{1}{n}}(\sin\gamma)^{-N-1}d\sigma(y)}$$

$$+ c_3 n^{-1}\sqrt{\int_{r_0 \leq \gamma \leq \pi - \frac{1}{n}}d\sigma(y)} \leq c_1 n^{\frac{N-1}{2}-\alpha}.$$

For the estimation of $I_2$ we use (3) of Lemma 2.4. We have

$$I_2 \leq c_4 n^{N-1-\alpha} \sqrt{\int_{\pi-\frac{1}{n}\leq \gamma \leq \pi} d\sigma(y)} \leq Cn^{N-1-\alpha} n^{-\frac{N}{2}} = Cn^{\frac{N}{2}-1-\alpha}.$$

Finally the norm of the Riesz means of the spectral function of the Laplace-Beltrami operator can be estimated as follows:

$$\left\|\Theta^\alpha(x,y,n)\right\|_{L_2(S^N\setminus V)} \leq I_1 + I_2 \leq Cn^{\frac{N-1}{2}-\alpha}.$$

Lemma 2.3 is proved.

We denote

$$\Theta^\alpha_{\frac{l}{2}}(x,y,n) = \overline{(1-\Delta_s)}^{\frac{l}{2}} \Theta^\alpha(x,y,n)$$

Using the summation by parts we have

$$\Theta^\alpha_{\frac{l}{2}}(x,y,n) = \sum_{k=0}^{n}\left(1-\frac{\lambda_k}{\lambda_n}\right)^\alpha (1+\lambda_k)^{\frac{l}{2}} \sum_{j=1}^{a_k} Y_j^k(x) Y_j^k(y)$$

$$= \sum_{k=0}^{n-1}\left[(1+\lambda_k)^{\frac{l}{2}} - (1+\lambda_{k+1})^{\frac{l}{2}}\right]\Theta^\alpha(x,y,k) + (1+\lambda_n)^{\frac{l}{2}}\Theta^\alpha(x,y,n).$$

The norm of $\Theta^\alpha_{\frac{l}{2}}(x,y,n)$ can be estimated as follows

$$\left\|\Theta^\alpha_{\frac{l}{2}}(x,y,n)\right\|_{L_2} \leq \sum_{k=0}^{n-1}\left|(1+\lambda_k)^{\frac{l}{2}} - (1+\lambda_{k+1})^{\frac{l}{2}}\right|\left\|\Theta^\alpha(x,y,k)\right\|_{L_2}$$

$$+(1+\lambda_n)^{\frac{l}{2}}\left\|\Theta^\alpha(x,y,n)\right\|_{L_2}. \tag{4}$$

For the second summand we have

$$(1+\lambda_n)^{\frac{l}{2}}\left\|\Theta^\alpha(x,y,n)\right\|_{L_2^l(S^N\setminus V)} \leq C(1+n)^{\frac{N-1}{2}-\alpha+l}.$$

Here we have used Lemma 2.4. For any $l > 0$ the following estimation holds

$$\left|(1+\lambda_k)^{\frac{l}{2}} - (1+\lambda_{k+1})^{\frac{l}{2}}\right| \leq C(1+k)^{l-1}, \quad k = 1, 2, ..., K$$

Applying the latter to the first summand in (4) gives

$$\sum_{k=0}^{n-1}\left[(1+\lambda_k)^{\frac{l}{2}} - (1+\lambda_{k+1})^{\frac{l}{2}}\right]\left\|\Theta_k^\alpha\right\|_{L_2^l(S^N \setminus V)} \leq \sum_{k=1}^{n-1} c_1(1+k)^{l-1} c_2(1+k)^{\frac{N-1}{2}-\alpha}.$$

Finally we obtain

$$\left\|\Theta_{\frac{l}{2}}^\alpha(x,y,n)\right\|_{L_2^l(S^N \setminus V)} \leq C n^{\frac{N-1}{2}-\alpha+l}.$$

**Lemma 2.5** Let $f \in H_2^{-l}(S^N)$, $l > 0$ and let $V$ be some domain in $S^N$. If $f = 0$ in $V$, then one has

$$\left|E_n^\alpha f(x)\right| \leq C n^{\frac{N-1}{2}-\alpha+l} \|f\|_{-l},$$

uniformly in any compact $K \subset V$.

**Proof.** From that $f$ vanishes in the domain $V \subset S^N$, and from (2) we obtain for any $u \in C^\infty(S^N)$

$$|<f, u>| \leq \|f\|_{-l} \|u\|_{H_2^l(S^N \setminus V)}. \tag{5}$$

Note that the operators $(\overline{1-\Delta_s})^{-\frac{l}{2}}$, $l > 0$ are continuous from $L_2$ to $H_2^l$, i.e for any $u \in L_2$ we have,

$$\|u\|_{H_2^l} = \left\|(\overline{1-\Delta_s})^{-\frac{l}{2}} u\right\|_{L_2} \leq \|u\|_{L_2}. \tag{6}$$

Taking into account (5) and (6) one has

$$\left|E_n^\alpha f(x)\right| = \left|<f, \Theta^\alpha(x,y,n)>\right| \leq \|f\|_{-l} \left\|\Theta^\alpha(x,y,n)\right\|_{H_2^l(S^N \setminus V)}$$

$$\|f\|_{-l} \left\|\Theta_{\frac{l}{2}}^\alpha(x,y,n)\right\|_{L_2(S^N \setminus V)} \leq C(1+n)^{\frac{N-1}{2}-\alpha+1} \|f\|_{-l},$$

where $\alpha \geq \frac{N-1}{2} + l.$

**Proof of Theorem 2.2.**

Let $f \in H_2^{-l}(S^N), l > 0$ and $f$ coincides with $g$ in $V \subset S^N$, where $g(x)$ is continuous function on $(S^N)$. For all functions $g \in C(S^N)$ Riesz means $E_n^\alpha f(x)$ uniformly converges to $g$ on any compact $K \subset S^N$.

Let consider a new distribution $F(x) = f(x) - g(x), x \in V$ and $F(x) = f(x), x \in S^N \setminus V$. It is obvious that $F(x) \in H^{-l}(S^N)$ and vanishes on $V$. Then for any compact $K \subset V$ we have

$$\left|E_n^\alpha F(x)\right| \leq C(1+n)^{\frac{N-1}{2}-\alpha+1} \|F\|_{-l}.$$

By choosing $\alpha, l$, such that $\alpha > \frac{N-1}{2} + l$ we can make $E_n^\alpha F(x) \to 0$, as $n \to \infty$.

The representation

$$E_n^\alpha f(x) = E_n^\alpha F(x) + E_n^\alpha g(x)$$

and the fact that $E_n^\alpha g(x)$ uniformly converges to $g(x)$ on any compact $K \subset V$ completes the proof of Theorem 2.2.

## ACKNOWLEDGEMENT

This paper has been supported by Ministry of Higher Education (Mohe) for its MyPhd sponshorship and Universiti Putra Malaysia under Research University Grant (RUGS), project number 05-01-09-0674RU.